\documentstyle[fleqn,psfig,epsf]{elsart}

\begin{document}

\newcommand \ga{\raisebox{-.5ex}{$\stackrel{>}{\sim}$}}
\newcommand \la{\raisebox{-.5ex}{$\stackrel{<}{\sim}$}}
\newcommand \prc{Phys. Rev. C}
\newcommand \prd{Phys. Rev. D}
\newcommand \prl{Phys. Rev. Lett.}
\newcommand \apjl{Ap.J.Lett.}
\newcommand \apj{Ap.J.}
\newcommand \nphysa{nucl. phys. A}
\newcommand \physrep{Phys. Rep.}
\newcommand \bea{\begin{eqnarray}}
\newcommand \eea{\end{eqnarray}}

\begin{frontmatter}

\title{PHASES OF DENSE MATTER IN NEUTRON STARS}

\author{Henning Heiselberg}

\address{NORDITA, Blegdamsvej 17, DK-2100 Copenhagen \O, Denmark}

\maketitle

\begin{abstract}

After a brief history of neutron stars and supernovae recent
developments are discussed.  Based on modern nucleon-nucleon potentials
more reliable equations of state for dense nuclear matter have been
constructed.  Furthermore, phase transitions such as pion, kaon and
hyperon condensation, superfluidity and quark matter can occur in
cores of neutron stars. Specifically, the nuclear to quark matter
phase transition and its mixed phases with intriguing structures is
treated.  Rotating neutron stars with and without phase transitions
are discussed and compared to observed masses, radii and glitches.
The observations of possible heavy $\sim 2M_\odot$ neutron stars in
X-ray binaries and QPO's require relatively stiff equation of states
and restricts strong phase transitions to occur at very high nuclear
densities only.

\end{abstract}

\end{frontmatter}

\newpage

\section{INTRODUCTION AND HISTORY OF NEUTRON STARS}

 It is a great pleasure to participate in the 6'th Colloque Cosmologie
at the Observatoire de Paris and stand on the same historic spot, where
my fellow countryman Ole R\o mer more than two centuries 
ago first calculated the speed of 
light\footnote{
Actually, he measured and (over-)estimated the light time from the sun, 
$\sim 12min$, but did not bother to calculate the speed of light because
the very distance to the sun was not well known at that time.}.
I shall therefore start with a brief history of neutron stars and
subsequently discuss some very important and recent developments
in this field at the turn of this millenium.
 The most important discoveries concerning neutron stars are
briefly listed in Table I. Most are well known except perhaps for
the most recent ones, which will be discussed in more detail below.
\begin{table}[htp]
\begin{center}
\caption{Chronological list of important developments related to 
neutron stars}         
\begin{tabular}{rll} \\\hline\noalign{\smallskip}
     Year & ``Observers''    & Discovery \\
\noalign{\smallskip}\hline\noalign{\smallskip}
     1054 & Chinese          & record the Crab Supernova   \\
     1572 & Tycho Brahe      & observes a Supernova    \\
     1604 & Kepler           & observes a Supernova     \\
     1932 & Chadwick         & discovers the neutron  \\
     1932 & Landau           & suggests the existence of neutron stars\\
     1934 & Baade \& Zwicky  & connects supernovae to gravitational\\
          &                  &  collapse of stars to neutron stars\\
     1935 & Oppenheimer \& Volkoff  & calculate
            neutron star structures with simple EoS  \\
%     1946 & Gamow            & developes nucleosynthesis which later\\
%          &                  & requires heavier elements from Supernovae\\
     1967 & Bell \& Hewish   & discover first pulsar  \\
     1969 &          & pulsars in Crab and Vela Supernova remnants \\
     1973 & Hulse \& Taylor  & discover first binary pulsar  \\
     1987 & Neutrino detectors  & collects 19 neutrinos from SN-1987A  \\
     1995 & Nijmegen data base  & compilation of $\ga5000$ NN cross sections\\
          &                  & leads to ``modern'' NN potentials and EoS\\
     1996 & RXTE  & discover kHz oscillations (QPO) in X-ray binaries \\
     1997 & BeppoSAX & Gamma Ray Burst with afterglow at $z\ga1$ \\
\noalign{\smallskip}\hline
\label{history}
\end{tabular}
\end{center}
\end{table}

\section{HEAVY NEUTRON STARS IN X-RAY BINARIES}

The best determined neutron star masses are found in binary pulsars
and all lie in the range $1.35\pm 0.04 M_\odot$ (see Thorsett and
Chakrabarty 1999). These masses have been accurately determined from
variations in their radio pulses due to doppler shifts as well
periastron advances of their close elliptic orbits that are
strongly affected by general relativistic effects.
One exception is the nonrelativistic pulsar PSR J1012+5307
of mass\footnote{The uncertainties are all 95\% conf. limits or $\sim2\sigma$}
$M=(2.1\pm 0.8)M_\odot$ (van Paradijs 1998). 

Several X-ray binary
masses have been measured of which the heaviest are 
Vela X-1 with $M=(1.9\pm 0.2)M_\odot$ (Barziv et al., 1999)
and Cygnus X-2 with
$M=(1.8\pm 0.4)M_\odot$ (Orosz \& Kuulkers 1999).  
Their Kepler orbits are determined by measuring doppler shift of
both the X-ray binary and its companion. To complete the mass
determination one needs the orbital inclination which is determined
by eclipse durations, optical light curves, or polarization variations
(see, e.g., van Paradijs).

The recent
discovery of high-frequency brightness oscillations in low-mass X-ray
binaries provides a promising new method for determining masses and
radii of neutron stars (see Miller, Lamb, \& Psaltis 1998). The
kilohertz quasi-periodic oscillations (QPO) occur in pairs
and are most likely the orbital
frequencies 
\bea
   \nu_{QPO}=(1/2\pi)\sqrt{GM/R_{orb}^3} \,, \label{orb}
\eea
of accreting matter
in Keplerian orbits around neutron stars of mass $M$ and its beat
frequency with the neutron star spin, $\nu_{QPO}-\nu_s$.  According to
Zhang, Strohmayer, \& Swank 1997, Kaaret, Ford, \& Chen (1997) the accretion
can for a few QPO's be tracked to its innermost stable orbit,
\bea
   R_{ms} &=& 6GM/c^2 \,. \label{iso}
\eea
 For slowly rotating stars the resulting mass is from 
Eqs. (\ref{orb},\ref{iso})
\bea
    M &\simeq& 2.2M_\odot \frac{{\mathrm{kHz}}}{\nu_{QPO}}  \,.
\eea
  For example, the
maximum frequency of 1060 Hz upper QPO observed in 4U 1820-30 gives $M\simeq
2.25M_\odot$ after correcting for the $\nu_s\simeq275$ Hz neutron star
rotation frequency.  If the maximum QPO frequencies of 4U 1608-52
($\nu_{QPO}=1125$ Hz) and 4U 1636-536 ($\nu_{QPO}=1228$ Hz) also
correspond to innermost stable orbits the corresponding masses are
$2.1M_\odot$ and $1.9M_\odot$.  
Evidence for the innermost stable orbit has only been found
for 4U 1820-30, where $\nu_{QPO}$ display a distinct saturation
with accretion rate indicating that orbital frequency cannot 
exceed that of the innermost stable orbit.
More observations like that are needed before a firm conclusion can
be made.

Such large neutron star masses of order $\sim 2M_\odot$ severely restrict
the equation of state (EoS) for dense matter as addressed in the
following.

\section{MODERN NUCLEAR EQUATION OF STATES}

Recent models for the nucleon-nucleon (NN) interaction, based on the
compilation of more than 5000 NN cross sections in the Nijmegen data
bank, have reduced the uncertainty in NN potentials.  Including
many-body effects, three-body forces, relativistic effects, etc., the
nuclear EoS have been constructed with reduced uncertainty allowing
for more reliable calculations of neutron star properties, see Akmal,
Pandharipande, \& Ravenhall (1998) and Engvik et al.~(1997).
Likewise, recent realistic effective interactions for nuclear matter
obeying causality at high densities, constrain the EoS
severely and thus also the maximum masses of neutron stars, see Akmal,
Pandharipande, \& Ravenhall (1998) and Kalogera \& Baym (1996). We
have in \cite{physrep} elaborated on these analyses by incorporating causality
smoothly in the EoS for nuclear matter allowing for first and second
order phase transitions to, e.g., quark matter.

For the discussion of the gross properties of neutron stars we will
use the optimal EoS of Akmal, Pandharipande, \& Ravenhall (1998) 
(specifically the Argonne $V18 + \delta v +$ UIX$^*$ model- hereafter
APR98), which is based on the most recent models
for the nucleon-nucleon interaction, see Engvik et al.~(1997) for a
discussion of these models, and with the inclusion of a parametrized
three-body force and relativistic boost corrections. The EoS for
nuclear matter is thus known to some accuracy for densities up to a
few times nuclear saturation density $n_0=0.16$ fm$^{-3}$.  We
parametrize the APR98 EoS by a simple form for the compressional and
symmetry energies that gives a good fit
around nuclear saturation densities and smoothly incorporates
causality at high densities such that the sound speed approaches the
speed of light.
This requires that the compressional part of the
energy per nucleon is quadratic
in nuclear density with a minimum at saturation but linear at high densities
\begin{eqnarray}
    {\cal E} &=& E_{comp}(n) + S(n)(1-2x)^2 \nonumber\\
   &=& {\cal E}_0 u\frac{u-2-s}{1+s u} +S_0 u^\gamma (1-2x)^2.   
    \label{eq:EA} 
\end{eqnarray}
Here, $n=n_p+n_n$ is the total baryon density, $x=n_p/n$ the
proton fraction and $u=n/n_0$ is the ratio of the baryon density to
nuclear saturation density. The compressional term is in Eq.\
(\ref{eq:EA}) parametrized by a simple form which reproduces the
saturation density and the binding energy per
nucleon ${\cal E}_0=15.8$MeV at $n_0$ of APR98. The ``softness''
parameter $s\simeq 0.2$, which gave the best fit to the data
of APR98 (see Heiselberg \& Hjorth-Jensen 1999)  
is determined by fitting the energy per
nucleon of APR98 up to densities of $n\sim 4n_0$.
For the symmetry energy
term we obtain $S_0=32$ MeV and $\gamma=0.6$ for the best fit. The
proton fraction is given by $\beta$-equilibrium at a given density.

The one unknown parameter $s$ 
expresses the uncertainty in the EoS at high
density and we shall vary this parameter within the allowed limits in
the following with and without phase transitions to calculate mass,
radius and density relations for neutron stars.
The ``softness'' 
parameter $s$ is related to the incompressibility
of nuclear matter as $K_0=18{\cal E}_0/(1+s)\simeq 200$MeV. It 
agrees with the poorly known experimental value
(Blaizot, Berger, Decharge, \& Girod 1995), 
$K_0\simeq 180-250$MeV which does not restrict it
very well.  From $(v_s/c)^2=\partial P/\partial (n\cal{E})$, where $P$ is the
pressure, and the EoS  of Eq.\ (\ref{eq:EA}),
the causality condition $c_s\le c$ requires 
\begin{equation}
      s \ga \sqrt{\frac{{\cal E}_0}{m_n}} \simeq 0.13 \,,\label{causal}
\end{equation}
where $m_n$ is the mass of the nucleon.
With this condition we have a causal EoS that reproduces the
data of APR98 at densities up to $0.6\sim 0.7$ fm$^{-3}$. 
In contrast, the EoS of APR98 becomes
superluminal at $n\approx 1.1$ fm$^{-3}$.  For larger $s$ values
the EoS is softer which eventually leads to smaller maximum masses of
neutron stars. The observed $M\simeq 1.4M_\odot$ in binary pulsars
restricts $s$ to be less than $0.4-0.5$ depending on rotation
as shown in calculations of neutron stars
below. 

\begin{figure}\begin{center}
{\centering\mbox{\psfig{file=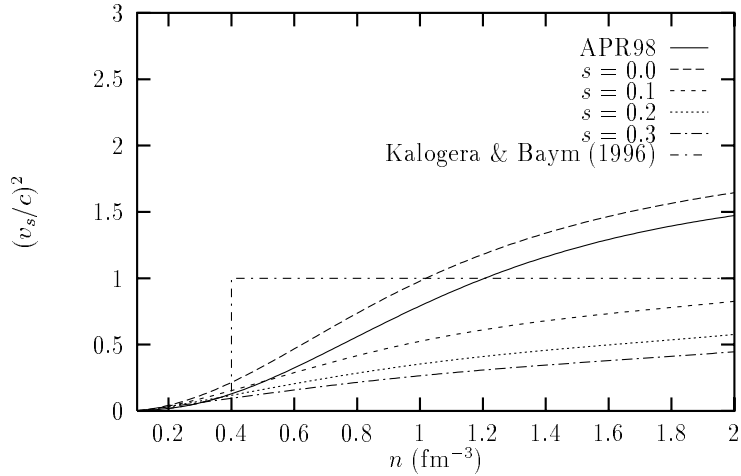,height=24cm,angle=0}}}
\vspace{-7cm}
       \caption{$(v_s/c)^2$ for $\beta$-stable $pn$-matter 
                 for $s=0,0.1,0.2,0.3$, 
                the results of APR98, 
                and for the 
                patched EoS of Kalogera \& Baym (1996) which shows 
                a discontinuous $(v_s/c)^2$. }
       \label{fig1}
\end{center}\end{figure}

In Fig.\ \ref{fig1} we plot the sound speed $(v_s/c)^2$ for
various values of $s$ and that resulting from the microscopic
calculation of APR98 for $\beta$-stable $pn$-matter.  The form of
Eq.~(\ref{eq:EA}), with the inclusion of the parameter $s$, provides
a smooth extrapolation from small to large densities
such that the sound speed $v_s$ approaches the
speed of light. For $s=0.0$ ($s=0.1$) the EoS
becomes superluminal at densities of the order of 1 (6) fm$^{-3}$.

The sound speed of Kalogera \& Baym (1996) is also plotted in Fig.\
\ref{fig1}. It jumps discontinuously to the speed of light at a
chosen density. With this prescription they were able to obtain an
optimum upper bound for neutron star masses and obey causality.  This
prescription was also employed by APR98, see Rhoades \& Ruffini (1974)
for further details.  The EoS is thus discontinuously stiffened by
taking $v_s=c$ at densities above a certain value $n_c$ which,
however, is lower than $n_{s}=5n_0$ where their nuclear EoS becomes
superluminal. This approach stiffens the nuclear EoS for densities
$n_c<n<n_s$ but softens it at higher densities. Their resulting
maximum masses lie in the range $2.2M_\odot\la M\la 2.9M_\odot$.  Our
approach however, incorporates causality by reducing the sound speed
smoothly towards the speed of light at high densities. Therefore our
approach will not yield an absolute upper bound on the maximum mass
of a neutron star
but gives reasonable estimates based on modern EoS around nuclear matter
densities, causality constraints at high densities and a smooth
extrapolation between these two limits (see Fig. 1).

At very high densities particles are expected to be relativistic 
and the sound speed should be smaller than the speed of light,
$v_s^2\simeq c^2/3$. Consequently, the EoS should be even softer at
high densities and the maximum masses we obtain with the EoS
of (\ref{eq:EA}) are likely to be too high estimates.

\section{PHASE TRANSITIONS}

 The physical state of matter in the interiors of neutron stars at
densities above a few times normal nuclear matter densities is
essentially unknown and many first and second order phase transitions
have been speculated upon. We will specifically study the hadron to
quark matter transition at high densities, but note that other
transitions as, e.g., kaon and/or pion condensation or the presence
of other baryons like hyperons also
soften the EoS and thus further aggravate the resulting reduction in
maximum masses. Hyperons appear at densities typically of the order
$2 n_0$ and result in a considerable softening of the EoS, see
e.g., Balberg, Lichenstadt, \& Cook (1998). Typically, most equations 
of state with
hyperons yield masses around $1.4-1.6 M_{\odot}$.   
Here however, in order to focus on  the role played by phase transitions
in neutron star matter, we will assume that a phase transition
from nucleonic to quark matter takes place at a certain density.
Since we do not have a fully reliable theory for the quark matter
phase, we will for simplicity employ the bag model in our
actual studies of quark phases and neutron star properties.  In
the bag model the quarks are assumed to be confined to
a finite region of space, the so-called 'bag', by a vacuum pressure
$B$.  Adding the
Fermi pressure and interactions computed to order $\alpha_s=g^2/4\pi$,
where $g$ is the QCD coupling constant, the total pressure
for 3 massless quarks of flavor $f=u,d,s$, is (see Kapusta (1988), 
\begin{equation}
    P=\frac{3\mu_f^4}{4\pi^2}(1-\frac{2}{\pi}\alpha_s) -B +P_e+P_\mu \,,
     \label{pquark}
\end{equation}
where $P_{e,\mu}$ are the electron and muon pressure, e.g., 
$P_e=\mu_e^4/12\pi^2$.
A Fermi gas of quarks of flavor {\em i} has density $n_i =
k_{Fi}^3/\pi^2$, due to the three color states. 
A finite strange quark mass have minor effect on the EoS since
quark chemical potentials $\mu_q\ga m_N/3$ typically are
much larger. The value of the bag constant {\em B} is poorly
known, and we present results using two representative values,
$B=150$ MeVfm$^{-3}$ and $B=200$ MeVfm$^{-3}$.
We take $\alpha_s=0.4$. However, similar results can be obtained with
smaller $\alpha_s$ and larger $B$  (Madsen 1998).

\begin{figure}
\begin{center}
{\centering\mbox{\psfig{file=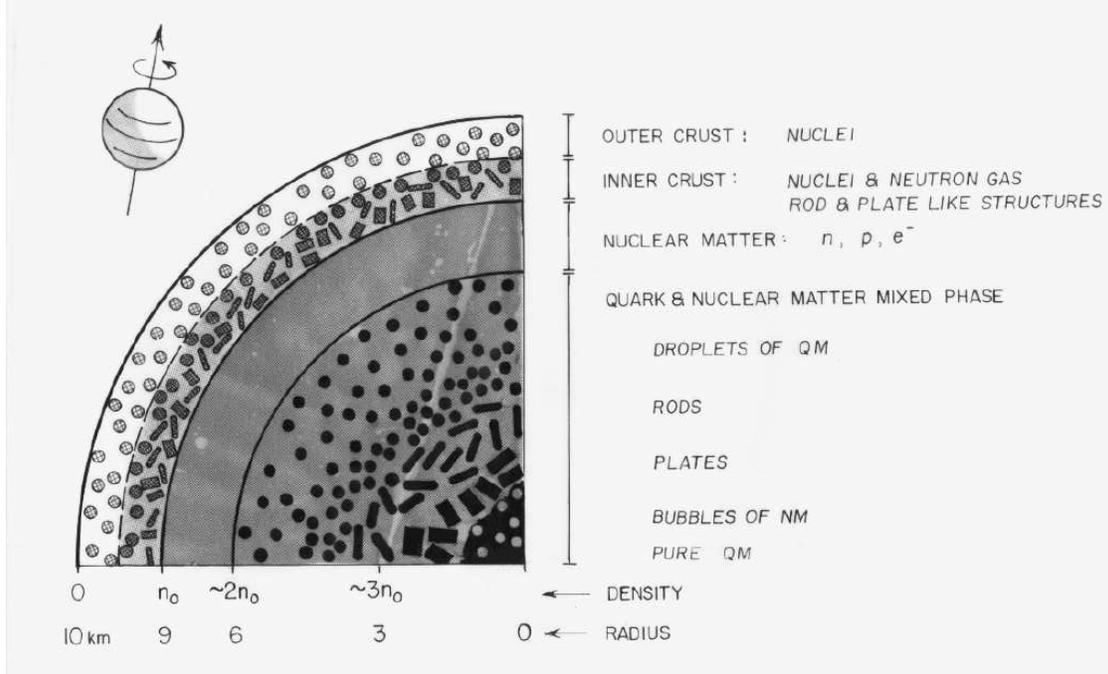,height=9cm,angle=90}}}
\vspace{1cm}
\caption{The quark and nuclear matter structure in a quarter of a typical
1.4$M_\odot$ solar mass neutron star.
The typical sizes of structures are a few Fermi's but have been scaled
up by about 16 orders of magnitudes to be seen.
}
\end{center}
\label{star}
\end{figure}

 The quark and nuclear matter mixed phase described in Glendenning
(1992) has continuous pressures and densities due to the general Gibbs
criteria for two-component systems. There are no first order phase
transitions but at most two second order phase transitions. Namely, at
a lower density, where quark matter first appears in nuclear matter,
and at a very high density (if gravitationally stable), where all
nucleons are finally dissolved into quark matter. This mixed phase
does, however, not include local surface and Coulomb energies of the
quark and nuclear matter structures. If the interface tension between
quark and nuclear matter is too large, the mixed phase is not favored
energetically due to surface and Coulomb energies associated with
forming these structures (Heiselberg, Pethick, \& Staubo 1993). The
neutron star will then have a core of pure quark matter with a mantle
of nuclear matter surrounding it and the two phases are coexisting by
a first order phase transition or Maxwell construction.
For a small or moderate interface tension the quarks are confined in
droplet, rod- and plate-like structures as found in the inner crust of
neutron stars (Lorenz, Ravenhall, \& Pethick 1993).

\section{MASSES AND RADII OF NEUTRON STARS}

In order to obtain the mass and radius of a neutron star, we have solved the
Tolman-Oppenheimer-Volkov equation with and without rotational
corrections following the approach of Hartle (1967).  The equations of
state employed are given by the $pn$-matter EoS with $s =0.13,
0.2, 0.3, 0.4$ with nucleonic degrees of freedom only. In addition we
have selected two representative values for the bag-model parameter
$B$, namely 150 and 200 MeVfm$^{-3}$ for our discussion on eventual phase
transitions. The quark phase is linked with our $pn$-matter EoS from
Eq.\ (\ref{eq:EA}) with $s=0.2$ through either a mixed phase
construction or a Maxwell construction, see Heiselberg and
Hjorth-Jensen (1999) for further details.  For $B=150$ MeVfm$^{-3}$, the
mixed phase begins at 0.51 fm$^{-3}$ and the pure quark matter phase
begins at $1.89$ fm$^{-3}$.  Finally, for $B=200$ MeVfm$^{-3}$, the mixed
phase starts at $0.72$ fm$^{-3}$ while the pure quark phase starts
at $2.11$ fm$^{-3}$.  In case of a Maxwell construction, in order to
link the $pn$ and the quark matter EoS, we obtain for $B=150$ MeVfm$^{-3}$
that the pure $pn$ phase ends at $0.92$ fm$^{-3}$ and that the
pure quark phase starts at $1.215$ fm$^{-3}$, while the corresponding
numbers for $B=200$ MeVfm$^{-3}$ are $1.04$ and $1.57$
fm$^{-3}$.

\begin{figure}[htp]
{\centering\mbox{\psfig{file=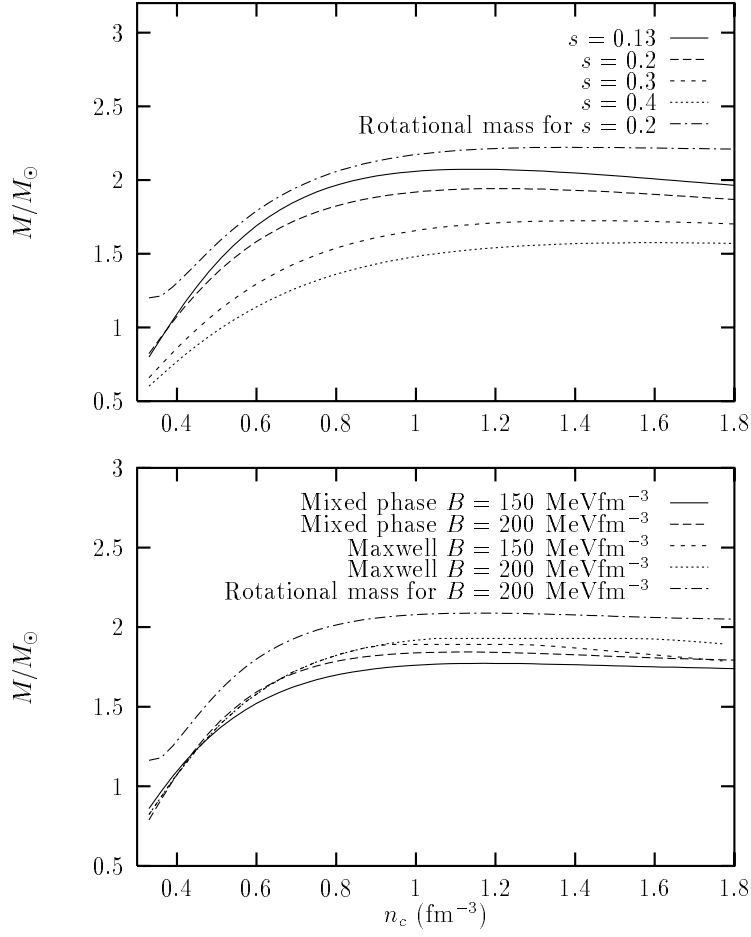,height=24cm,angle=0}}}
\vspace{-6cm}
\caption{Total mass $M$ as function of central density $n_c$
         for various values of $s$ (upper panel) 
         and the bag parameter $B$ (lower panel) for
         both a mixed phase and a Maxwell constructed EoS with 
         $s=0.2$ in Eq.~\ref{eq:EA}). 
         In addition we include also the rotational corrections
         for the pure $pn$-case with $s=0.2$ and the mixed
         phase construction for $B=200$ MeVfm$^{-3}$. For the Maxwell
         construction which exhibits a first order phase transition, 
         in the density regions where the two phases coexist, 
         the pressure is constant, a fact reflected
         in the constant value of the neutron star mass. All results
         are for $\beta$-stable matter. Note also that for the upper
         panel, the EoS for $s=0.3$ and $s=0.4$ start to differ
        from those with $s=0.13,0.2$ at densities below $0.2$ fm$^{-3}$.}
\label{fig2}
\end{figure}

None of the equations of state
from either the pure $pn$ phase or with a mixed phase or Maxwell
construction with quark degrees of freedom, result in stable
configurations for densities above $\sim 10 n_0$, implying thereby
that none of the stars have cores with a pure quark phase.  The EoS
with $pn$ degrees of freedom have masses $M\la2.2M_{\odot}$ when
rotational corrections are accounted for.  With the inclusion of the
mixed phase, the total mass is reduced since the EoS is softer.
 However, there is the possibility of
making very heavy quark stars for very small bag constants. For pure
quark stars there is only one energy scale namely $B$ which provides
a homology transformation (Madsen 1998) and the maximum mass is
$M_{max}=2.0M_\odot (58{\rm MeV fm^{-3}}/B)^{1/2}$ (for
$\alpha_s=0$). However, for $B\ga 58{\rm MeV fm^{-3}}$ a nuclear
matter mantle has to be added and for $B\la 58{\rm MeV fm^{-3}}$ quark
matter has lower energy per baryon than $^{56}$Fe and is thus the
ground state of strongly interacting matter. Unless the latter is the
case, we can thus exclude the existence of $2.2-2.3M_\odot$ quark
stars.

\begin{figure}
\begin{center}
{\centering\mbox{\psfig{file=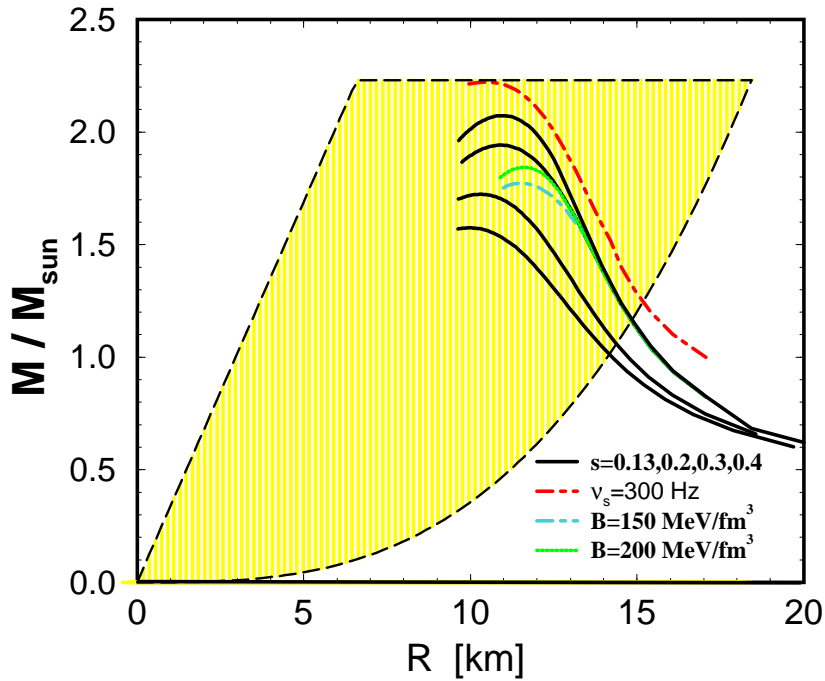,height=100mm,angle=0}}}
\caption{Neutron star masses vs.~radius for the EoS of Eq.~(1) 
with softness s=0.13,0.2,0.3,04, with increasing values of $s$ 
from top to bottom for the full curves.
Phase transitions decrease the maximum mass whereas rotation
increases it. The shaded area represents  the neutron star
radii and masses allowed (see text and Eqs. 1-3) for 
orbital QPO frequencies 1060~Hz of 4U 1820-30. 
%From Ref. \cite{HJ}
}
\end{center}
\label{fig4}
\end{figure}

In Fig.\ \ref{fig4} we show the mass-radius relations for the various
equations of state. 
The shaded area represents the allowed masses and radii for 
$\nu_{QPO}=1060$ Hz of 4U 1820-30. Generally,
\begin{eqnarray}
  2GM < R < \left(\frac{GM}{4\pi^2\nu_{QPO}^2}\right)^{1/3} \,,
\end{eqnarray}
where the lower limit insures that the star is not a black hole,
and the upper limit that the accreting matter orbits outside
the star, $R<R_{orb}$. Furthermore,
for the matter to be outside the innermost
stable orbit, $R>R_{ms}=6GM$, requires that 
\begin{eqnarray}
   M &\la& \frac{1+0.75j}{12\sqrt{6}\pi G\nu_{QPO}} \,
 \simeq 2.2 M_\odot (1+0.75j)\frac{{\rm kHz}}{\nu_{QPO}} \,,  \label{Mms} 
\end{eqnarray}
where $j=2\pi c\nu_sI/M^2$ is a dimensionless measure of the angular
momentum of the star with moment of inertia $I$.  The upper limit
in Eq. (\ref{Mms}) is the mass when $\nu_{QPO}$ corresponds to the innermost
stable orbit. According to Zhang, Smale, Strohmayer \& Swank (1998)
this is the case for 4U 1820-30 since
$\nu_{QPO}$ saturates at $\sim1060$~Hz with increasing count rate.
The corresponding neutron star mass is $M\sim 2.2-2.3M_\odot$ which
leads to several interesting conclusions as seen in Fig.\
\ref{fig4}. Firstly, the stiffest EoS allowed by causality
($s\simeq 0.13-0.2$) is needed. Secondly, rotation must be included
which increase the maximum mass and corresponding
radii by 10-15\% for $\nu_s\sim 300$~Hz.
Thirdly, a phase transition to quark matter below densities of order
$\sim 5 n_0$ can be excluded, corresponding to a restriction on the bag
constant $B\ga200$ MeVfm$^{-3}$.

 These maximum masses are smaller than those of APR98 and Kalogera \&
Baym (1996) who, as discussed above, obtain upper bounds on the mass
of neutron stars by discontinuously setting the sound speed to equal
the speed of light above a certain density, $n_c$. By varying the density
$n_c=2\to 5n_0$ the
maximum mass drops from $2.9\to 2.2M_\odot$. In our case,
incorporating causality smoothly by introducing the parameter $s$ in
Eq.\ (\ref{eq:EA}), the EoS is softened at higher densities in order
to obey causality, and yields a maximum mass which instead is slightly
lower than the $2.2M_\odot$ derived in APR98 for nonrotating stars.

If the QPOs are not from the innermost stable orbits and one finds
that even accreting neutron stars have small masses, say like the
binary pulsars, $M\la1.4M_\odot$, this may indicate that heavier
neutron stars are not stable. Therefore, the EoS is soft at high
densities $s\ga0.4$ or that a phase transition occurs at a few
times nuclear matter densities. For the nuclear to quark matter
transition this would require $B<80$ MeVfm$^{-3}$ for
$s=0.2$. For such small bag parameters there is an appreciable
quark and nuclear matter mixed phase in the neutron star interior but
even in these extreme cases a pure quark matter core is not obtained
for stable neutron star configurations.

\section{GLITCHES}

 Younger pulsars rotate and slow down rapidly. Some display
sudden speed ups referred to as glitches. We shall first
discuss such standard glitches and subsequently continue to
giant glitches and other characteristic glitch behavior for
spin frequencies close to the critical ones, where first order
phase transitions occur at densities present right at the core of neutron
stars.

\subsection{Core quakes and glitches} \label{subsec:glitches}

The glitches observed in the Crab, Vela, and a few other pulsars are
probably due to quakes occurring in solid structures such as the
crust, superfluid vortices or possibly the quark matter lattice in the
core \cite{HPS}.  As the rotating neutron star gradually slows down and
becomes less deformed, the rigid component is strained and eventually
cracks/quakes and changes its structure towards being more spherical.

The moment of inertia of the rigid component, $I_c$, decreases
abruptly and its rotation and pulsar frequency increases due to
angular momentum conservation resulting in a glitch.  The observed
glitches are very small $\Delta\Omega/\Omega\sim 10^{-8}$.  The two
components slowly relaxate to a common rotational frequency on a
time scale of days (healing time) due to superfluidity of the other
component (the neutron liquid).  The {\it healing parameter}
$Q=I_c/I_{tot}$ measured in glitches reveals that for the Vela and
Crab pulsar about $\sim$3\% and $\sim$96\% of the moment of inertia is
in the rigid component respectively.

If the crust were the only rigid component the Vela neutron star
should be almost all crust.  This would require that the Vela is a
very light neutron star - much smaller than the observed ones which
all are compatible with $\sim 1.4M_\odot$.  If we by the lattice
component include not only the solid crust but also the protons in 
nuclear matter (NM)
(which is locked to the crust due to magnetic fields), superfluid
vortices pinned to the crust \cite{Pines} and the solid QM mixed phase
\begin{equation}
   I_c = I_{crust}+I_p+I_{sv}+I_{QM} \, ,
\end{equation}
we can better explain the large $I_c$ for the Crab.
The moment of inertia of the mixed phase is sensitive to the EoS's used.
For example, for a quadratic NM EoS \cite{HPS} 
decreasing the Bag constant from 110 to 95 MeVfm$^{-3}$ increases
$I_c/I_{total}$ from $\sim20\%$ to $\sim70\%$ 
for a 1.4$M_\odot$ neutron star - not including possible vortex pinning.
The structures in the mixed phase would exhibit anisotropic elastic properties,
being rigid to some shear strains but not others in much the same way as liquid
crystals. Therefore the whole mixed phase might not be rigid.

The energy released in glitches every few years
are too large to be stored in the crust only. 
The recurrence time for large quakes, $t_c$, is inversely
proportional to the strain energy \cite{Pines}, 
which again is proportional to the lattice density and the Coulomb energy
\begin{equation}
   t_c^{-1} \propto  \frac{1}{a^3} \frac{Z^2e^2}{a} \, .
\end{equation}
Since the lattice distance $a$ is smaller for the quark matter droplets 
and their charge larger 
than for atoms in the crust, the recurrence
time is shorter in better agreement with measurements of large glitches.

Detecting core and crust quakes
separately or other signs of three components in glitches, indicating
the existence of a crust, superfluid neutrons and a solid core, would
support the idea of the mixed quark and nuclear matter mixed
phase. However, magnetic field attenuation is expected to be small in
neutron stars and therefore magnetic fields penetrate through the
core. Thus the crust and core lattices as well as the proton liquid
should be strongly coupled and glitch simultaneously.

\subsection{Backbending and giant glitches}

In \cite{gpw97} the moment of inertia is found to ``backbend'' as
function of angular velocity.  The moment of inertia of some deformed
nuclei \cite{MV,Johnson} may also backbend when the coriolis force
exceeds the pairing force breaking the pairing whereby the nucleus
reverts from partial superfluidity to a rigid rotor. However, in the
limit of large nuclear mass number such backbending would
disappear. Instead pairing may lead to superfluidity in bulk
\cite{Pines}.  A backbending phenomenon  in neutron stars,
that appears to be similar to backbending in nuclei, can occur in
neutron stars although the physics behind is entirely different.
If we soften the EoS significantly at a density near the central density
of the neutron star, a non-rotating neutron star can have most of
its core at high densities where the soft EoS determines the
profile. A rapidly rotating star may instead have lower central
densities only probing the hard part of the EoS. Thus the star may at
a certain angular velocity revert from the dense phase to a more
dilute one and at the same time change its structure and moment of
inertia discontinuosly. Such a drastic change in moment of inertia
at some angular velocity will cause a giant glitch as found in \cite{gpw97}.

 The phenomenon of neutron star backbending is related to the
double maximum mass for a neutron and quark star respectively as discussed
in Ref. \cite{physrep}. Both phenomena occur when the ``softening''
or density discontinuity exceeds
\begin{eqnarray}
  \varepsilon_2 \ga \frac{3}{2} \varepsilon_1 
 \quad \Rightarrow \quad {\rm Two\, maximum\, masses\, and\,
Giant \, glitch\, when\, \omega\simeq\omega_0}. 
 \label{ecrito}
\end{eqnarray}
Similarly, there is also a
discontinuity in the moment of inertia when $\rho_2/\rho_1\ga
3/2$. The neutron star may continue to slow down in its unstable
structure, i.e. ``super-rotate'', before reverting to its stable
configuration with a dense core.  As for the instabilities of Eq.\
(\ref{ecrito}) this condition changes slightly for a more general EoS and
when general relativity is included.  Neutron stars with a mixed phase
do not have a first order phase transition but may soften their EoS
sufficiently that a similar phenomenon occurs \cite{gpw97}.  We
emphasize, however, that the discontinuous jump in moment of inertia
is due to the drastic and sudden softening of the EoS near the central
density of neutron stars.  It is not important whether it is a phase
transition or another phenomenon that causes the softening.

\subsection{Phase transitions in rotating neutron stars} \label{subsec:PTR}

During the last years, and as discussed in the previous sections,
interesting phase transitions in nuclear matter 
to quark matter, 
mixed phases of quark and nuclear matter \cite{Glendenning,HPS}, kaon
\cite{kn87} or pion condensates \cite{Akmal}, 
neutron and proton superfluidity \cite{eeho96},
hyperonic matter \cite{schulze97,sm96,kpe96}
crystalline nuclear matter, magnetized matter, etc., have
been considered.  Here we consider the
interesting phenomenon of how the star and in particular its moment
of inertia behaves near the critical angular velocity, $\Omega_0$,
where the core
pressure just exceeds that needed to make a phase transition. 

\begin{figure}
\begin{center}
{\centering
\mbox{\psfig{figure=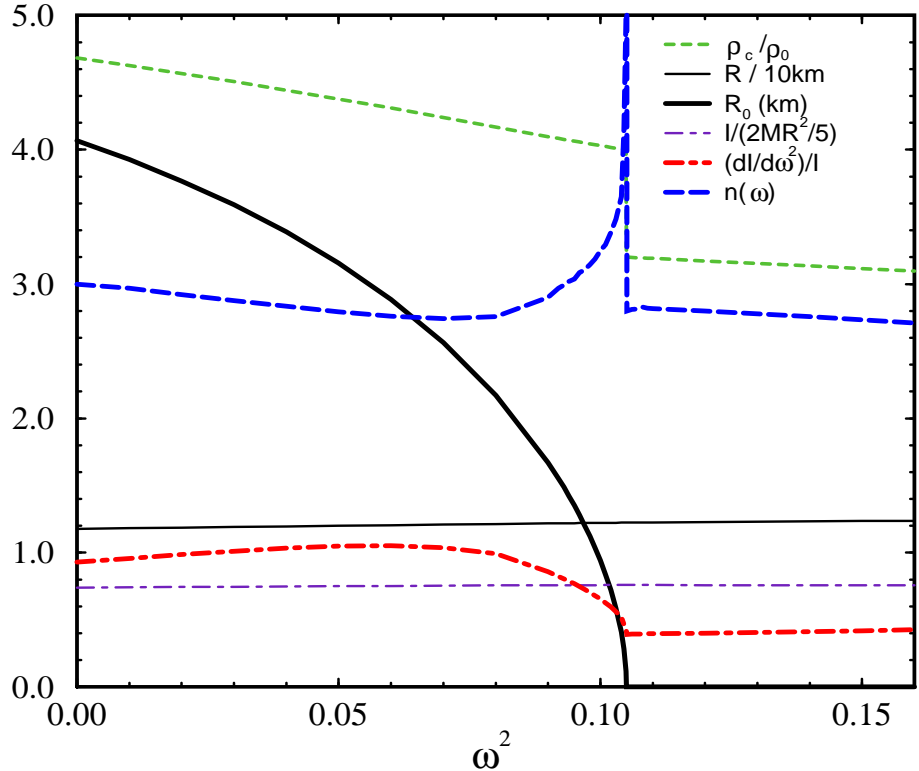,height=12cm,width=15cm,angle=-90}} }
\caption{Central density (in units of $\varepsilon_0$), radii of the neutron
star $R$ and its dense core $R_0$, moment of inertia, its derivative
$I'/I=dI/d\omega^2/I$ and the braking index are shown as function of
the scaled angular velocity $\omega^2=\Omega^2/(2\pi \varepsilon_1)$.
The rotating neutron star has mass $1.4M_\odot$ and a Bethe-Johnson
like polytropic EoS with a first order phase transition
taking place at density $\varepsilon_1=3.2\varepsilon_0$ to
$\varepsilon_2=4\varepsilon_0$. }       \label{rot}
\end{center}
\end{figure}

As described in detail in \cite{physrep} the general relativistic 
equations for slowly rotating stars can be solved even with
first order phase transitions. The resulting moment of inertia
have the characteristic behavior for 
$\Omega \raisebox{-.5ex}{$\stackrel{<}{\sim}$}\Omega_0$
(see also Fig. (\ref{rot}))
\begin{equation}
  I = I_0\left( 1+\frac{1}{2}c_1\frac{\Omega^2}{\Omega_0^2} -\frac{2}{3}c_2 
                (1-\frac{\Omega^2}{\Omega_0^2})^{3/2} + ...
      \right) . 
  \label{Igen}
\end{equation}
For the two incompressible fluids with momentum of inertia, 
the small expansion
parameters are $c_1=\omega_0^2$ and 
$c_2=(5/2)\omega_0^3(\varepsilon_2/\varepsilon_1-1)/(3-2\varepsilon_2/
\varepsilon_1-\omega_0^2)^{3/2}$; for $\Omega>\Omega_0$ the $c_2$ term is 
absent. 
For a Bethe-Johnson polytropic EoS we find from Fig.\ (\ref{rot}) that
$c_2\simeq 0.07\simeq 2.2\omega_0^3$. Generally, we find that the
coefficient $c_2$ is proportional to the density difference between the 
two coexisting phases and to the critical angular velocity to the third power,
$c_2\sim (\varepsilon_2/\varepsilon_1-1)\omega_0^3$. The scaled critical 
angular velocity
$\omega_0$ can at most reach unity for submillisecond pulsars.

To make contact with observation we consider the temporal behavior
of angular velocities of pulsars. The pulsars slow down at a rate
given by the loss of rotational energy which we shall assume is
proportional to the rotational angular velocity to some power
(for dipole radiation $n=3$)
\begin{equation}
  \frac{d}{dt} \left(\frac{1}{2}I\Omega^2\right) = -C \Omega^{n+1}. 
   \label{dE}
\end{equation}
With the moment of inertia given by Eq. (\ref{Igen})
the angular velocity will then decrease with time as
\begin{eqnarray}
  \frac{\dot{\Omega}}{\Omega} &=& -\frac{C\Omega^{n-1}}{I_0}
  \left( 1-c_1\frac{\Omega^2}{\Omega_0^2}
          -c_2\sqrt{1-\frac{\Omega^2}{\Omega_0^2}} \right) \nonumber\\
  &\simeq& -\frac{1}{(n-1)t}
  \left( 1-c_2 \sqrt{1-(\frac{t_0}{t})^{2/(n-1)}} +....\right), 
  \label{dOmega}
\end{eqnarray}
for $t\ge t_0$. Here,
the time after formation of the pulsar is, using Eq.\  (\ref{dE}),
related to the angular velocity as
$t\simeq t_0(\Omega_0/\Omega)^{n-1}$ and 
$t_0=I_0/((n-1)C\Omega_0^{n-1})$ for $n>1$, is the critical time where
a phase transition occurs in the center. For earlier times $t\le t_0$ there
is no dense core and Eq. (\ref{dOmega}) applies when setting $c_2=0$
The critical angular velocity is $\Omega_0=\omega_0\sqrt{2\pi 
\varepsilon_1}\simeq
6 kHz$ for the Bethe-Johnson EoS \cite{BJ}, 
i.e.\  comparable to a millisecond binary
pulsar. Applying these numbers to, for example, the Crab pulsar we find that
it would have been spinning with critical angular velocity approximately 
a decade after the Crab supernova explosion,
i.e.\  $t_0\sim 10$ years for the Crab. Generally,
$t_0\propto\Omega_0^{1-n}$ and the timescale for the transients in
$\dot{\Omega}$ as given by Eq. (13) may be months or centuries. In any
case it would not require continuous monitoring which would help
a dedicated observational program.

The braking index depends on the second derivative $I''=dI/d^2\Omega$
of the moment of inertia and thus diverges (see Fig. (1))
as $\Omega$ approaches $\Omega_0$ from below
\begin{eqnarray}
     n(\Omega) &\equiv& \frac{\ddot{\Omega}\Omega}{\dot{\Omega}^2} 
    \simeq n - 2c_1\frac{\Omega^2}{\Omega_0^2}
    +c_2\frac{\Omega^4/\Omega_0^4}{\sqrt{1-\Omega^2/\Omega_0^2}} \,.\label{n}
\end{eqnarray}
For $\Omega\ge\Omega_0$ the term with $c_2$ is absent.
The  {\it observational} braking index $n(\Omega)$ should be distinguished
from the {\it theoretical} exponent $n$ appearing in Eq. (12).
Although the results in Eqs. (13) and (14) were derived 
for the pulsar slow down assumed in Eq. (12) both
$\dot{\Omega}$ and $n(\Omega)$ will generally 
display the $\sqrt{t-t_0}$ behavior for $t\ge  t_0$ as long as the rotational
energy loss is a smooth function of $\Omega$.
The singular behavior will, however, be smeared on the pulsar
glitch ``healing'' time which in the case of the Crab pulsar is 
of order weeks only.

We now discuss possible phase transitions in interiors of neutron
stars.  The quark and nuclear matter mixed phase described in
\cite{Glendenning} has continuous pressures and densities. There are
no first order phase transitions but at most two second order phase
transitions. Namely, at a lower density, where quark matter first
appears in nuclear matter, and at a very high density (if gravitationally
stable), where all nucleons are finally dissolved into quark matter.
In second-order
phase transitions the pressure is a continuous function of density
and we find a continuous braking
index.  This mixed phase does, however, not include local surface and
Coulomb energies of the quark and nuclear matter structures. As shown
in \cite{HPS} there can be an appreciable surface and Coulomb
energy associated with forming these structures and if the interface
tension between quark and nuclear matter is too large, the mixed phase
is not favored energetically. The neutron star will then have a core
of pure quark matter with a mantle of nuclear matter surrounding it and
the two phases are coexisting by a first order phase transition.  For
a small or moderate interface tension the quarks are confined in
droplet, rod- and plate-like structures as found in the
inner crust of neutron stars \cite{lrp93}. 
Due to the finite Coulomb and
surface energies associated with forming these structures, the
transitions change from second order to first order at each
topological change in structure. 
If a Kaon condensate
appears it may also have such structures \cite{Schaffner}.
Pion condensates \cite{pion}, crystalline nuclear matter \cite{Akmal},
hyperonic or magnetized matter, etc. may provide
other first order phase transitions.

Thus, if a first order phase transitions is present at central
densities of neutron stars, it will show up in moments of inertia and
consequently also in angular velocities in a characteristic way.  For
example, the slow down of the angular velocity has a characteristic
behavior $\dot{\Omega}\sim c_2\sqrt{1-t/t_0}$ and the braking index
diverges as $n(\Omega)\sim c_2/\sqrt{1-\Omega^2/\Omega_0^2}$ (see
Eqs. (\ref{dOmega}) and (\ref{n})).  The magnitude of the signal
generally depends on the density difference between the two phases and
the critical angular velocity $\omega_0=\Omega_0/\sqrt{2\pi
\varepsilon_1}$ such that
$c_2\sim(\varepsilon_2/\varepsilon_1-1)\omega_0^3$.  The observational
consequences depend very much on the critical angular velocity
$\Omega_0$, which depends on the EoS employed, at which
density the phase transition occurs and the mass of the neutron star.

We encourage a dedicated search for the characteristic
transients discussed above. As the pulsar slows down over a 
million years, its central densities spans a wide range of order
$1 n_0$ (see Fig. \ref{rot}). As we are interested in time scales
of years, we must instead study all $\sim 1000$ pulsars available.
By studying the corresponding 
range of angular velocities for the sample of different
star masses, the chance for encountering a critical angular velocity
increases.  Eventually, one may be able to cover the
full range of central densities and find all first order phase
transitions up to a certain size determined by the experimental
resolution.  Since the size of the signal scales with $\Omega_0^3$ the
transition may be best observed in rapidly rotating pulsars such as
binary pulsars or pulsars recently formed in supernova explosion and
which are rapidly slowing down. Carefully monitoring such pulsars may
reveal the characteristic behavior of the angular velocity or braking
index as described above which is a signal of a first order phase
transition in dense matter.

\section{SUMMARY}

Modern nucleon-nucleon potentials have reduced the uncertainties in
the calculated EoS.  Using the most recent realistic effective
interactions for nuclear matter of APR98 with a smooth extrapolation
to high densities including causality, the EoS could be
constrained by a ``softness'' parameter $s$ which parametrizes the
unknown stiffness of the EoS at high densities. Maximum masses have
subsequently been calculated for rotating neutron stars with and
without first and second order phase transitions to, e.g., quark
matter at high densities.

The calculated bounds for maximum masses leaves two natural options
when compared to the observed neutron star masses:

\begin{itemize}
 \item {\bf Case I}: {\it The large masses of the neutron stars in
QPO 4U 1820-30 ($M=2.3M_\odot$), PSR J1012+5307
($M=2.1\pm0.4 M_\odot$), Vela X-1 ($M=1.9\pm0.1 M_\odot$), and
Cygnus X-2 ($M=1.8\pm0.2 M_\odot$), are confirmed and
complemented by other neutron stars with masses around $\sim 2M_\odot$.}

As a consequence, the EoS of dense nuclear matter is severely
restricted and only the stiffest EoS consistent with causality are allowed,
i.e., softness parameter $0.13\le s\la0.2$.  Furthermore, any
significant phase transition at densities below $< 5n_0$ can be
excluded. 

 That the radio binary pulsars all have masses around $1.4M_\odot$ is
then probably due to the formation mechanism in supernovae where the
Chandrasekhar mass for iron cores are $\sim1.5M_\odot$.
Neutron stars in binaries can subsequently acquire larger
masses by accretion as X-ray binaries.

 \item {\bf Case II}: 
{\it The heavy neutron stars prove erroneous by more detailed observations
and only masses like those of binary pulsars are found.}

If accretion does not produce neutron stars heavier than 
$\ga1.4M_\odot$, this indicates that heavier neutron stars simply are not
stable which in turn implies a soft EoS, either $s> 0.4$ or a
significant phase transition must occur already at a few times nuclear
saturation densities. 

\end{itemize}

Surface temperatures can be estimated from spectra and from the
measured fluxes and known distances, one can extract the surface
area of the emitting spot. This gives unfortunately only a lower limit
on the neutron star size, $R$. If it becomes possible to measure
both mass and radii of neutron stars, one can plot an observational
$(M,R)$ curve in Fig. (\ref{fig4}), which uniquely determines the
EoS for strongly interacting matter at zero temperature.

Pulsar rotation frequencies and glitches are other promising signals
that could reveal phase transitions. Besides the standard glitches
also giant glitches were mentioned and in particular the
characteristic behavior of angular velocities when a first order
phase transition occurs right in the center of the star.

It is, unfortunately, impossible to cover all the interesting
and recent developments concerning neutron stars in these proceedings
so for more details we refer to \cite{physrep} and Refs. therein.

{\bf Acknowledgements:} to my collaborators M. Hjorth-Jensen
and C.J.~Pethick as well as G. Baym, F. Lamb and V. Pandharipande.

\end{document}